\documentclass[acmtog, authorversion]{acmart} 
\acmSubmissionID{851}

\usepackage{wrapfig}
\usepackage{xtab}

\AtBeginDocument{%
  \providecommand\BibTeX{{%
    \normalfont B\kern-0.5em{\scshape i\kern-0.25em b}\kern-0.8em\TeX}}}

\copyrightyear{2024}
\acmYear{2024}
\setcopyright{rightsretained}
\acmConference[SA Conference Papers '24]{SIGGRAPH Asia 2024 Conference
Papers}{December 3--6, 2024}{Tokyo, Japan}
\acmBooktitle{SIGGRAPH Asia 2024 Conference Papers (SA Conference Papers
'24), December 3--6, 2024, Tokyo, Japan}\acmDOI{10.1145/3680528.3687661}
\acmISBN{979-8-4007-1131-2/24/12}

\begin{document}

\title{Motion-Driven Neural Optimizer for Prophylactic Braces Made by Distributed Microstructures}

\author{Xingjian Han}
\authornotemark[1]
\orcid{0000-0002-1899-1952}
\affiliation{%
 \department{Computer Science}
  \institution{Boston University}
  \city{Boston}
  \country{USA / The University of Manchester, United Kingdom}
}
\email{xjhan@bu.edu}

\author{Yu Jiang}
\authornote{Equal contribution of the first two authors.}
\orcid{0000-0003-1579-7158}
\affiliation{%
  \institution{Dalian University of Technology}
  \city{Dalian}
  \country{China / The University of Manchester, United Kingdom}
}
\email{To be filled}

\author{Weiming Wang}
\orcid{0000-0001-6289-0094}
\affiliation{%
 \department{Department of Mechanical and Aerospace Engineering}
  \institution{The University of Manchester}
  \city{Manchester}
  \country{United Kingdom}
}
\email{weiming.wang@manchester.ac.uk}

\author{Guoxin Fang}
\orcid{0000-0001-8741-3227}
\affiliation{%
  \institution{The Chinese University of Hong Kong}
  \city{Hong Kong}
  \country{China}
}
\email{guoxinfang@cuhk.edu.hk}

\author{Simeon Gill}
\orcid{0000-0002-5719-7516}
\affiliation{%
  \institution{The University of Manchester}
  \city{Manchester}
  \country{United Kingdom}
}
\email{simeon.gill@manchester.ac.uk}

\author{Zhiqiang Zhang}
\orcid{0000-0003-0204-3867}
\affiliation{%
 \department{Electronic and Electrical Engineering}
  \institution{University of Leeds}
  \city{Leeds}
  \country{United Kingdom}
}
\email{z.zhang3@leeds.ac.uk}

\author{Shengfa Wang}
\orcid{0000-0001-9030-833X}
\affiliation{%
 \department{To be filled}
  \institution{Dalian University of Technology}
  \city{Dalian}
  \country{China}
}
\email{sfwang@dlut.edu.cn}

\author{Jun Saito}
\orcid{0000-0002-3128-9009}
\affiliation{%
  \institution{Adobe Research}
  \city{Seattle}
  \country{USA}
}
\email{jsaito@adobe.com}

\author{Deepak Kumar}
\orcid{0000-0003-1728-484X}
\affiliation{%
 \department{Physical Therapy}
  \institution{Boston University}
  \city{Boston}
  \country{USA}
}
\email{kumard@bu.edu}

\author{Zhongxuan Luo}
\orcid{0000-0001-5997-2646}
\affiliation{%
  \institution{Dalian University of Technology}
  \city{Dalian}
  \country{China}
}
\email{zxluo@dlut.edu.cn}

\author{Emily Whiting}
\orcid{0000-0001-7997-1675}
\affiliation{%
 \department{Computer Science}
  \institution{Boston University}
  \city{Boston}
  \country{USA}
}
\email{whiting@bu.edu}

\author{Charlie C. L. Wang}
\authornote {Corresponding author}
\orcid{0000-0003-4406-8480}
\affiliation{%
 \department{Department of Mechanical and Aerospace Engineering}
 \institution{The University of Manchester}
 \country{United Kingdom}
}
\email{changling.wang@manchester.ac.uk}

\renewcommand{\shortauthors}{Han, Jiang, Wang, Fang, Gill, Zhang, Wang, Saito, Kumar, Luo, Whiting, Wang}

\begin{abstract}
Joint injuries, and their long-term consequences, present a substantial global health burden. Wearable prophylactic braces are an attractive potential solution to reduce the incidence of joint injuries by limiting joint movements that are related to injury risk. Given human motion and ground reaction forces, we present a computational framework that enables the design of personalized braces by optimizing the distribution of microstructures and elasticity. As varied brace designs yield different reaction forces that influence kinematics and kinetics analysis outcomes, the optimization process is formulated as a differentiable end-to-end pipeline in which the design domain of microstructure distribution is parameterized onto a neural network. The optimized distribution of microstructures is obtained via a self-learning process to determine the network coefficients according to a carefully designed set of losses and the integrated biomechanical and physical analyses. Since knees and ankles are the most commonly injured joints, we demonstrate the effectiveness of our pipeline by designing, fabricating, and testing prophylactic braces for the knee and ankle to prevent potentially harmful joint movements. 
\end{abstract}



\begin{teaserfigure}
  \includegraphics[width=\textwidth]{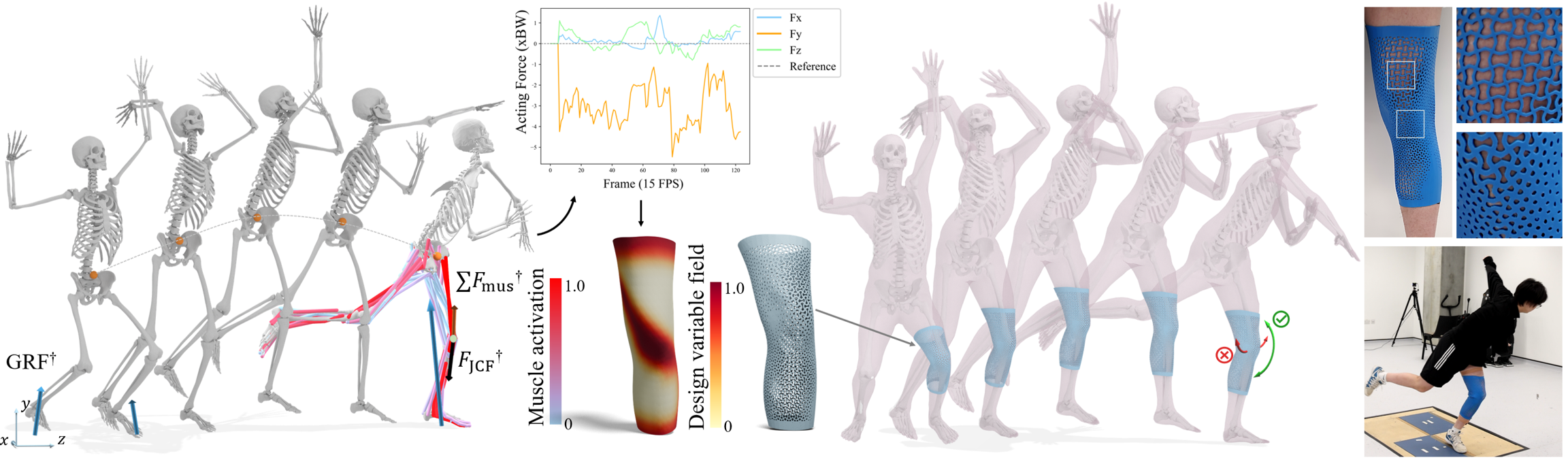}
\caption{
Given an input motion paired with ground reaction forces (GRFs), our approach produces a controlled distribution of elasticity physically realized by microstructures for prophylactic braces. The time-variant load given by kinetic analysis is plotted in the middle, where the frames have been downsampled to 15 fps and the acting forces are normalized to body weight (the testing participant weighs 64 $kg$). With the help of the optimized brace design, the frontal angles can be effectively reduced while keeping the flexibility of motion throughout the range within sagittal plane. 
}
\small\textsuperscript{$\dagger$} The force vectors include the GRF, the joint-contact force and the muscle forces are all scaled down by $0.5 \times$ for visualization purposes.
\Description{Teaser}
\label{fig:teaser}
\end{teaserfigure}

\maketitle
\section{Introduction}
In the realm of sports and recreational activities, the potential for injury presents a universal challenge. A well-designed prophylactic brace needs to provide thorough protection without constraining performance. Many current simulation-based design methods (e.g., \cite{ANDERSEN2023143}) employ a two-stage process by iteratively updating the input of the biomechanical analysis (i.e., the brace designs and the reaction forces) and the input loads of the physical analysis (i.e., the action forces). They require separate phases for initial development and subsequent integration along with manual iterations, which makes it challenging to benefit from the functionality of topology optimization for generating distributed elasticity. 
~\cite{LIU2023117190}. 
To overcome this limitation, we propose a computational pipeline that enables time-variant topology optimization equipped with kinematics and kinetics analysis (Fig.\ref{fig:teaser}). The effectiveness of our approach has been verified by physically fabricated braces through try-on tests and motion evaluations.

While more material may provide more support, distributing coverings over highly mobile regions might restrict mobility ~\cite{velko202210.1145/3526113.3545674}. Our design strategy aims for a balance: crafting a prophylactic brace that offers precise kinematics control of limiting undesired rotation (typically in the frontal plane) while preserving range of motion within the sagittal plane to execute movements safely. To achieve this, we consider both movement patterns (kinematics) and the forces involved (kinetics) in the design with integrated biomechanical and physical analyses.
We propose a design optimization for distributing elasticity, fabricated as microstructures in a hyperelastic material. The optimization is parameterized in a neural network with a self-learning process. We implement a carefully designed set of losses combined with integrated biomechanical and physical analyses. The result is a physical brace providing tailor-made prevention and preservation throughout the motion sequence. In summary, our contributions are as follows:
\begin{itemize}
    
    \item We investigate a closed-loop differentiable framework that can systematically optimize a customized brace design by integrated biomechanical analysis, physics analysis, and time-variant topology optimization (Sec.~\ref{sec:pipelineOverview}).
    
    \item We develope a formulation that effectively generates distributed microstructures based on the optimized neural function to achieve the desired elasticity distribution (Sec.~\ref{sec:FGM}).

\end{itemize}
We use our pipeline to design prophylactic braces and showcase the applicability of the method through designs with different joint dynamics and conditions.

\section{Related Work}

\subsection{Computational Design of Prophylactic Wearables}
Prophylactic bracing is a widely used and practical intervention for preventing the physical consequences of joint injuries. In particular, prophylactic knee braces are designed to control knee motion that is thought to increase the risk of anterior cruciate ligament (ACL) and other soft tissue injuries at the knee ~\cite{tuang2023biomechanical}. Reports \cite{Abrams492, majewski2006epidemiology} state that lateral collateral ligament and medial meniscus pathology were more frequent in tennis players as compared to other sports. ~\citet{tuang2023biomechanical} suggests reducing peak knee abduction (valgus) angles in the coronal (frontal) plane can potentially lower the risk of noncontact ACL injury. In the sagittal plane, promoting greater knee flexion can reduce the risk of ACL injury as landing with an extended knee with a reduced flexion angle range can increase the load on the ACL. ~\cite{Myer2011}. For ankle, prophylactic braces or taping aims to reduce the incidence of injuries like lateral sprains (e.g., ~\cite{Verhagen2010, Gross2003AnkleBracing, Farwell2013AnkleBraces, Hagan2024ProphylacticBraces}). These biomechanical insights guide our objective design of precise kinematics control over different planes to ensure joint stability.

Biomechanical simulation tools, such as AnyBody~\cite{ANDERSEN2023143} and OpenSim~\cite{opensim4352056}, offer a promising avenue for improving the design and optimization with the finite element analysis (FEA) for customization. We employ the biomechanical tools provided by OpenSim \cite{opensim4352056} to realize an end-to-end computational pipeline of motion-driven optimization.

In computational design, research has been conducted for decades to provide compression with customized shape of braces (e.g., \cite{Wang2007}). Recently, ~\citet{zhang2019customization} used topology optimization to design custom lightweight, high-performance compression casts/braces on two-manifold mesh surfaces, and \citet{XTZ10.1145/3126594.3126600} designed personalized orthopedic casts which are thermal-aware. \citet{jiang2022machine} demonstrated a customizable ankle brace with tunable stiffness by integrating machine learning. 
\citet{vechev2022computational_a} designed passively reinforced kinesthetic garments that resist a single motion and ~\citet{vechev2022computational} proposed an automatic optimization method to design connecting structures that efficiently counter a range of predefined body movements. However, these existing approaches have yet to consider the influence of reaction forces generated by varied designs of braces on the biomechanical model.

\begin{figure*}[t]
\centering
\setlength{\unitlength}{\linewidth}
  \begin{picture}(1,0.49)%
      \put(0,0){\includegraphics[width=\linewidth]{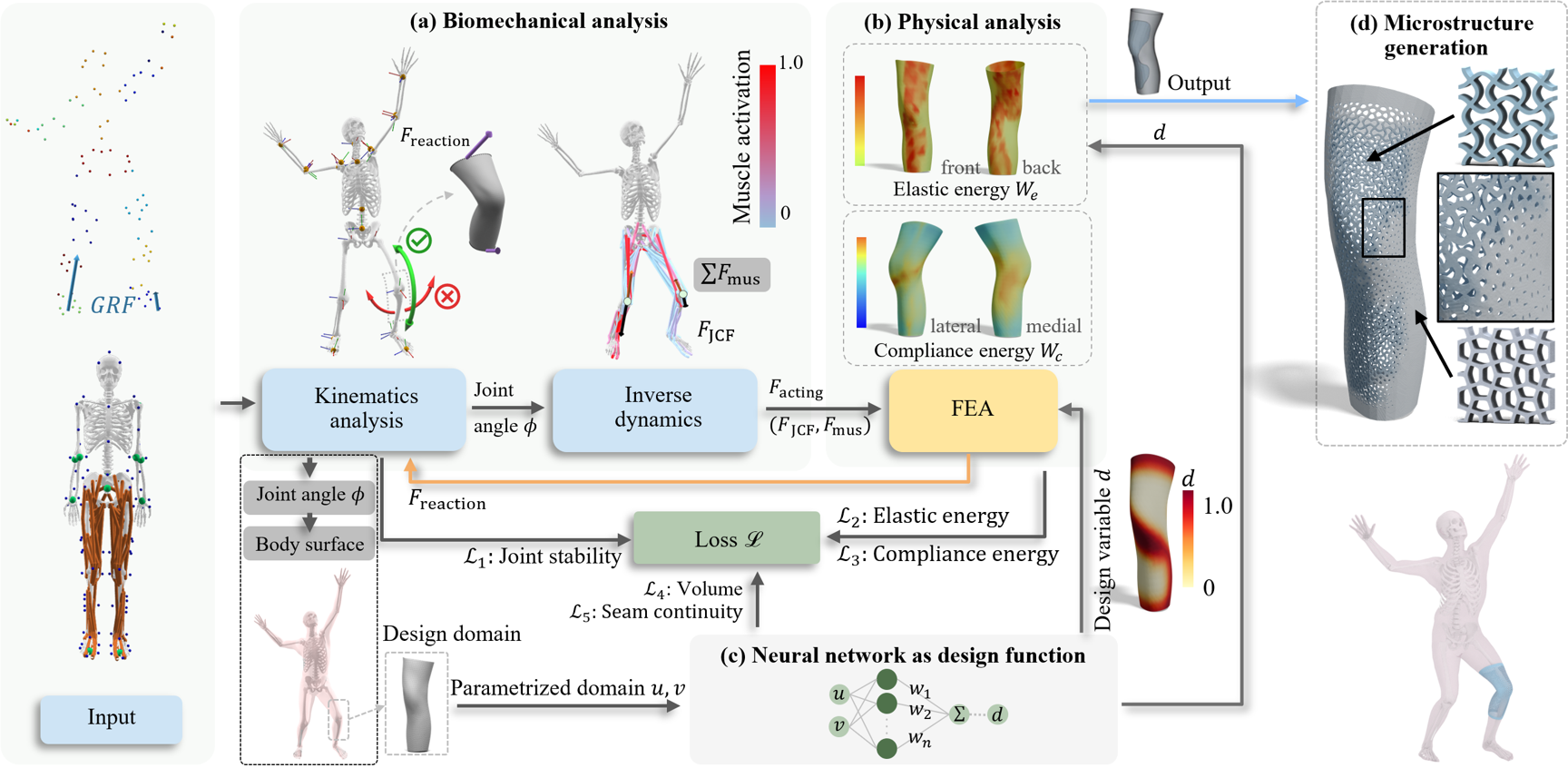}}%
    \small
  \end{picture}
 \caption{
Overview of our computational pipeline, which includes four major parts: (a) biomechanical analysis (Sec.\ref{sec:biomechAnalysis}), (b) physical analysis (Sec.\ref{sec:designFramework}), (c) the \textit{neural network} (NN) based representation of design function (Sec.\ref{sec:nerualrepresentation} and Sec.\ref{sec:network}), and (d) microstructure generation (Sec.\ref{sec:FGM}), along with a biomechanical data preprocessing (Sec.\ref{sec:dataProcess}, marked in black dotted box). From the input motion sequence and paired ground reaction forces (GRFs), we first conduct a preprocessing step to determine the joint angle, the muscle forces, and the design domain as the body surface with the help of a predefined musculoskeletal model. These terms remain unchanged throughout the optimization (as marked in gray blocks). Our topology optimization process, informed by the time-variant forces $F_\text{acting}$ given by the biomechanical analysis (a) and governed by the FEA-based physical analysis (b), iteratively updates the NN's weights (c) as design variables to change the distribution of stiffness on a brace while minimizing the compliance energy $W_c$. In our pipeline, the biomedical analysis and the physical analysis are closely coupled since the reaction force exerted by the braces will further impact the kinematics. The forces generated by a brace at its upper and lower boundaries $F_\text{reaction}$ are applied to the femur and tibia bones, respectively, prompting updates to the joint contact forces for subsequent analysis. (d) The distributed microstructures for realizing the optimized distribution of stiffness can be generated by first computing the distribution of elastic energy $W_e$ and then filling the regions in low and high elastic energies with the well-blended firm (displayed in light gray) and soft microstructures (displayed in light blue).
}\label{fig:pipeline}
\end{figure*}

\subsection{Topology Optimization}
Time-variant topology optimization is crucial for optimizing structures under time-dependent conditions. Unlike the traditional methods developed for static problems, only a few studies can be found in the literature on this topic (e.g., \cite{jensen2009space,wang2020space}). Their strategy of space-time extension is employed in our pipeline.

Topology optimization for microstructures was introduced by combining numerical homogenization in \citet{sigmund1994materials}. Many algorithms have since been developed to optimize the unit geometry of microstructures for desired performance, e.g. heterogeneous materials~\cite{torquato2010optimal}, functionally graded properties~\cite{radman2013topology}, and two-scale designs with multi-material microstructures~\cite{zhu2017two}. We employ a strategy similar to \citet{chen2018computational} to generate the distributed microstructures with Solid Isotropic Material with Penalization (SIMP) approach.

Recently, the computation of topology optimization has started to employ Neural Network (NN) techniques to leverage benefits such as auto-differentiation and highly parallel computing power. ~\citet{zhang2021tonr} and ~\citet{chandrasekhar2021tounn} directly optimized the density field alongside the update of NN's weights and bias, this method was recently extended to designing microstructure cells~\cite{sridhara2022generalized}. \citet{NeuMetMatNet10.1145/3618325} introduced a neural representation for generating structures based on directional stiffness and poisson ratio profiles. NN-based computation also offers functional and continuous representation of the design field, facilitating meshless analysis and optimization (e.g., \cite{zehnder2021ntopo}). Unlike deep neural networks methods, ~\citet{qian2023topology} uses a single layer of Gaussian activation functions, which capture nonlinear features, allowing to optimize fewer neurons (i.e., design variables) and enabling faster convergence. We adopt a similar network architecture.
\section{Motion-Driven Neural Optimizer}
\label{sec:pipelineOverview}
We introduce a topology design pipeline for prophylactic braces integrated with biomechanical analysis. Our pipeline produces an optimized elasticity distribution given time-variant loads acting on the human body. An overview can be found in Fig.~\ref{fig:pipeline}. A table of all used symbols can be found in the appendix (Table.~\ref{table:symbolfull}).

\subsection{Data Processing}
\label{sec:dataProcess}
\subsubsection{Inputs} 
\begin{figure}[t]
\centering
\setlength{\unitlength}{\linewidth}
  \begin{picture}(1,0.5)%
      \put(0,0){\includegraphics[width=\linewidth]{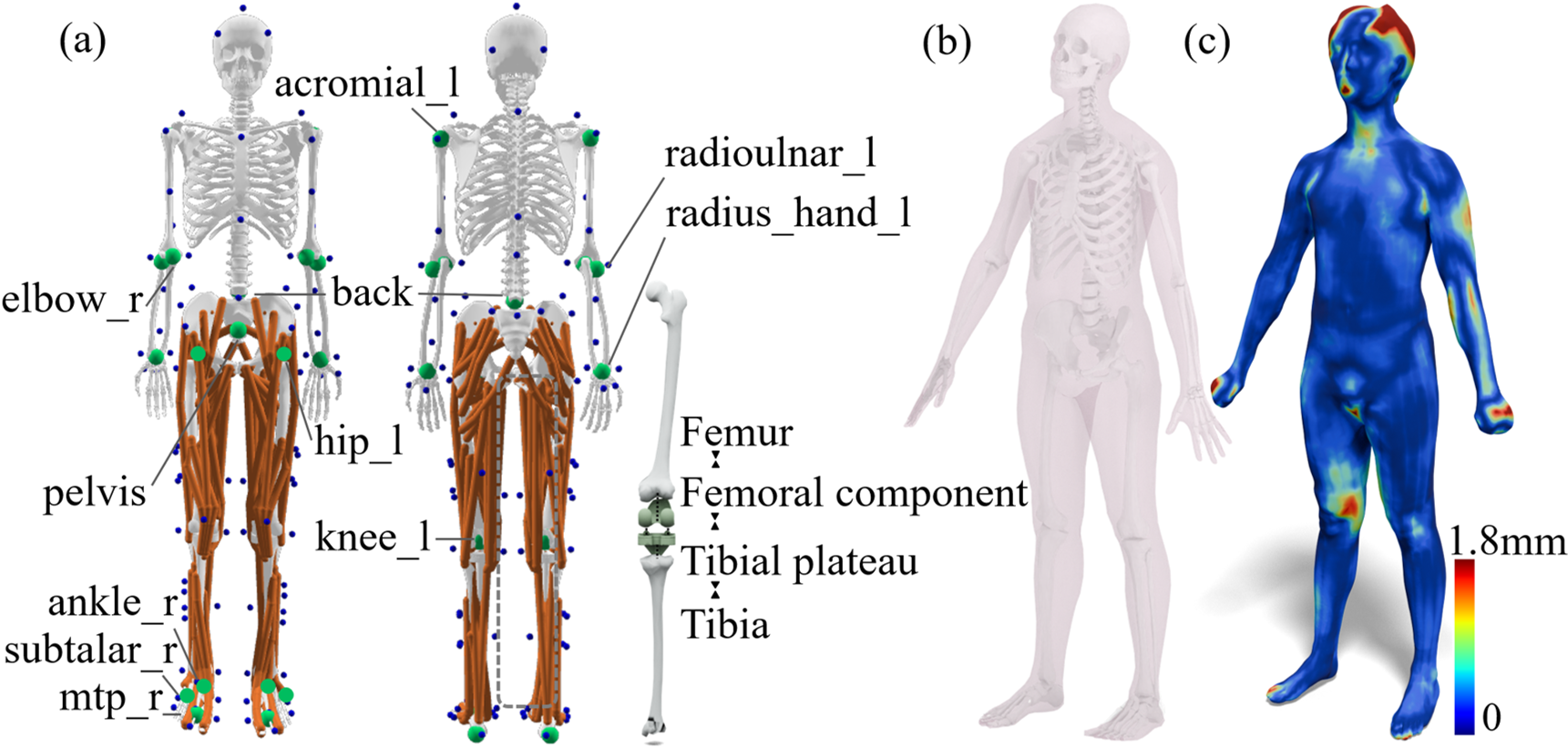}}%

    \small
  \end{picture}
\caption{
(a) Overview of the biomechanical model and the marker layout alignment for kinematics and kinetics analysis, where we customize a detailed knee compartment model ~\cite{biomodelLERNER2015644, kneeModelctx344555731410001631} by equipping both knees with three DoFs and adjusting the marker layout (with 96 markers as shown with blue dots) to accommodate more detailed kinematics. (b) From the musculoskeletal model, we extract the outer skin surface by fitting it to the SKEL model ($\beta$, $\mathbf{q}$)~\cite{keller2023skel}, where the shape parameters $\beta$ are obtained from the scanned body as shown in (c) to estimate detailed deformations. The color map measures the shape approximation error between the scanned and updated SKEL bodies.}
\label{fig:biomechModel}
\end{figure}

The paired motion capture (MoCap) data of 96 markers to describe full-body motion and corresponding ground reaction forces (GRF) are used as the input for defining movement and the external force resulting from the body's contact with the ground~\cite{han2023groundlink}. We customize a musculoskeletal model (Fig.~\ref{fig:biomechModel}(a)) from ~\cite{biomodelLERNER2015644,kneeModelctx344555731410001631} with $N=41$ degrees of freedom (DoF) including 3 DoFs for each knee. The model contains 80 Hill-type muscles to evaluate the muscle forces. We customize the locations of markers to ensure alignment with body positions, thereby enhancing the accuracy of the kinematic analysis. 

\subsubsection{Kinematics and optimization objectives}\label{sec:KinAnalysis}
From the MoCap and GRF data, we employ OpenSim~\cite{opensim4352056} and AddBiomechanics~\cite{AddBioWerling2023.06.15.545116} to calculate the joint angles from marker trajectories by inverse kinematics (IK), with $\mathbf{\phi}\in \Re^N$ representing the joint angles. 
Through optimization, we track joint angle changes in response to the additional reaction force exerted by the braces. We aim to restrict the movement in the frontal plane by minimizing the knee abduction/adduction and ankle inversion/eversion, as they can lead to joint instability and injuries like ACL tears or ankle sprain. Conversely, maintaining a full range of motion in the sagittal plane is essential to safeguard the human body, particularly during landing; these sagittal angles are the ones we aim to preserve. Note that different protocols and third-party software, Visual3D \cite{Visual3D}, are employed for results verification in Sec~\ref{sec:verification}.

\begin{figure}[t]
\centering
\setlength{\unitlength}{\linewidth}
  \begin{picture}(1,0.5)%
      \put(0,0){\includegraphics[width=\linewidth]{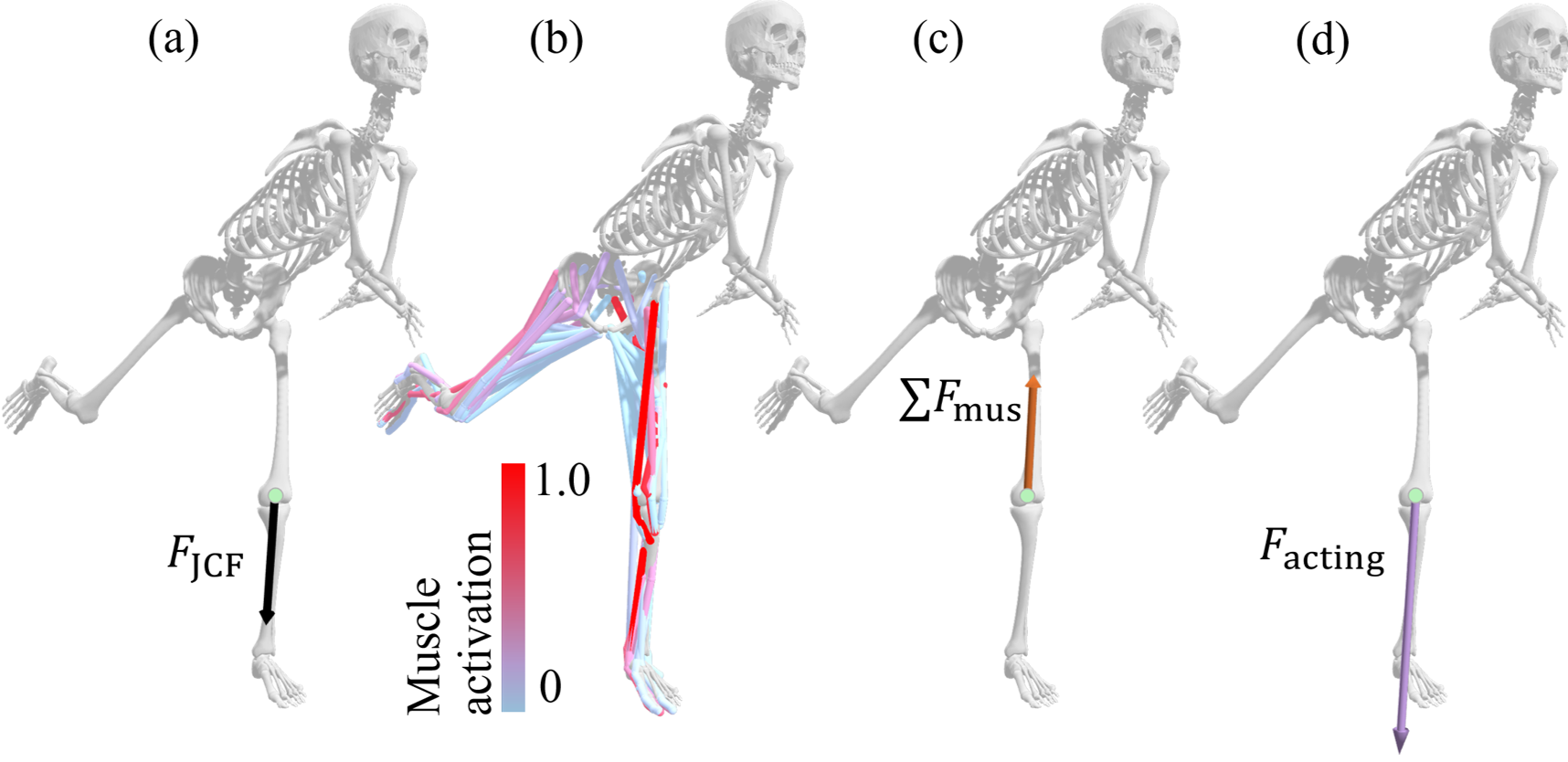}}%
    \small
  \end{picture}
\caption{We show the kinetic analysis of an example left knee joint during the tennis serve motion, specifically at the landing pose. The biomechanical dynamic analysis can estimate (a) the joint contact forces, $\mathbf{F}_{JCF}$, acting on the tibia bone, (b) the muscle activations and (c) the sum of muscle forces $\sum \mathbf{F}_{mus}$. The acting force (d) can then be computed as Eq.~(\ref{eq:actingForce}), which serves as the load input for the integrated FEA. Our optimizer considers the entire sequence of a complex motion, such as tennis serving, which includes several critical phases such as deep bending knees, jumping, and landing. 
}
\label{fig:kinetics}
\end{figure}

\subsubsection{Body reconstruction}
Extracting the outer skin surface directly from a musculoskeletal model is challenging due to complex anatomy and dynamic deformation introduced by motions. To address this, we employ the SKEL model $(\beta, \mathbf{q})$~\cite{keller2023skel}, which defines a parametric body model with corresponding skin and skeleton meshes. $\beta = [\beta_1, ... \beta_{10}]$ is the shape parameters, and $\mathbf{q}\in \mathbf{R}^{46}$ is the pose parameters representing 46 DoFs of the articulated body. We select some specific regions from the extracted body surface to construct the design domain (Fig.~\ref{fig:pipeline}). This allows us to analyze the impacts of motion / forces acting on the targeted areas.

\subsubsection{Parametrization for brace design} 

\label{sec:parametrization}
Customized braces can be simplified as two-manifold designs on a surface with uniform thickness. These designs are parameterized onto a domain as $(u,v) \in [0,1]$ using the least squares conformal maps ~\cite{levy2023least}. This reduces the dimensionality of the design domain, making it easier to conduct physical analysis and the generation of distributed microstructures. For taking FEA, the domain is tessellated into a triangular mesh $\mathcal{M}$. Note that the initial relaxed state of the body mesh instead of a deformed mesh is employed for parameterization in our implementation. Moreover, introducing a cutting line is necessary when parameterizing a brace surface resembling a cylinder onto a 2D plane. Constraints are added during the optimization to ensure the design variables at both sides of the cut are consistent. 

\subsection{Biomechanical Analysis}\label{sec:biomechAnalysis}

During human motion, forces and torques (or moments) acting across a joint are derived from external (ground reaction forces) and internal (body weight, muscle forces, soft tissue forces, joint contact forces) sources. At any given instant in time, the dynamic equilibrium is defined as $\sum\mathbf{F} = {m\mathbf{a}}$ and $\sum\mathbf{M} = {I \mathbf{\alpha}}$ with body segments' mass $m$, acceleration $\mathbf{a}$, inertia $I$, and angular acceleration $\mathbf{\alpha}$. Braces can generate reaction forces as additional external sources that alter the magnitude of other components of the dynamic equilibrium. 

\subsubsection{Joint contact forces}
A joint contact force ($\mathbf{F}_{\text{JCF}}$), as shown in Fig.~\ref{fig:kinetics}(a), represents the actual load transmitting between bones within a joint. While directly measuring the contact force is difficult, we model the contact with~\citet{opensim4352056} similar to ~\citet{biomodelLERNER2015644} and \citet{kneeModelctx344555731410001631} as shown in Fig.~\ref{fig:biomechModel}(a) to extract compressive force acting on the tibia bones for the knee and on the talus bones for the ankle joints. We compute $\mathbf{F}_{\text{JCF}}$ with \textit{joint reaction analysis} (JRA) ~\cite{opensim4352056} given the kinematics and the external loads defined by GRF and the brace reaction forces.

\subsubsection{Muscle forces} 
Muscles play a crucial role in supporting and stabilizing joints. A prophylactic brace can supplement muscle by sharing the load. Following ~\citet{opensim4352056} and ~\citet{feng2023musclevae}, we model the muscle as polylines. Muscle activation 
falling in the range of $[0,1]$ determines the level of contraction in a muscle, which is used to estimate the muscle forces. During the computation, we only consider the muscles traversing through the joint. The direction of the net muscle force is determined by considering each involved muscle line. We denote the sum of the muscle forces as $\sum \mathbf{F}_{\text{mus}}$.

\subsubsection{Overall acting force} The overall input forces for FEA are the net loads acting on the joint, accounting for both passive (due to the joint and body weight) and active (due to muscle support) components. Specifically, we define the force as follows, and the acting force serves as time-variant loads for FEA:

\begin{equation}
\begin{split}
 \mathbf{F}_{\text{acting}} = \mathbf{F}_{\text{JCF}}(\mathbf{F}_{\text{GRF}}, \mathbf{F}_{\text{reaction}}) - \sum \mathbf{F}_{\text{mus}} \\
\end{split}
\label{eq:actingForce}
\end{equation}
With these boundary conditions, we can formulate the optimization problem to compute the optimized topology for distributed elasticity.
\subsection{Motion-driven Topology Design}
Traditional topology optimization algorithms focus on spatial design domains, such as density fields. However, they may not be directly applicable or efficient for addressing time-variant loading problems arising from kinesthetic brace designs. This limitation stems from the high computational cost associated with variable-time nonlinear physical models and the non-convex nature of the design space ~\cite{CAI2023108355}. Developing or adapting algorithms to efficiently navigate this complex landscape remains challenging. Therefore, we integrate the space field with the motion frame, concurrently addressing them through discretization and optimization within a framework.

\label{sec:designFramework}
\subsubsection{Time-variant Topology Optimization Model}
\label{sec:nerualrepresentation}
By borrowing the concept of the density-based SIMP method ~\cite{sigmund200199}, we construct the time-variant topology optimization driven by action forces obtained from the kinematics and kinetics analysis. The design domain $d(u,v)$ (similar to `density' concept in SIMP) is represented by a neural network (NN), which is parameterized by the network coefficients $\theta$. For numerical computation, the SIMP-like optimization is conducted by tessellating the whole domain into finite elements. The design variable $d_e$ in each element $e$ is chosen by the field value at the center of $e$, where different values of $d_e$ will be mapped to different volume fractions and mechanical properties.
In motion-involved scenarios, loads and body surface can change with time; the time-variant loads can be represented as $\mathbf{F}(t)$ with a total of ${N_t}$ frames sampled at the rate of 15~fps as shown in Fig.\ref{fig:teaser}. The walking motion shown in Fig.~\ref{fig:designResults} is sampled at 60~fps.

The time-variant topology optimization problem can then be solved by minimizing the `worst' compliance energy as:

\begin{eqnarray}
  \label{eq:top}
  &\min_{d} \max_{\left \{ t_i \right \} }  &\left\{\frac{1}{2}\mathbf{U}^{T}\mathbf{K}(d)\mathbf{U}\right\}
  \\
  &{\rm s.t.} & \mathbf{K}(d)\mathbf{U}=\mathbf{F}_{\text{acting}}(t_i), \nonumber \sum_{e=1}^{N_l} \left(f_V(d_e)\cdot V_e\right) \leq V_0, \label{eq:volfrac} \nonumber 
\end{eqnarray}
with $e=1,2,...,N_l, i=1,2,...,N_t$, where $N_l$ is the number of elements, $f_E(.)$ and $f_V(.)$ are the functions of the Young’s modulus and the volume ratio related to the design function $d(\cdot)$, $\mathbf{K}$ and $\mathbf{U}$ are the global stiffness matrix and nodal displacement vector for FEA.
The FEA is conducted by planer elements equipped with the rotation matrix between 3D surface and its corresponding region in the $u,v$-domain. Specifically, we use MITC3 shell elements (Mixed Interpolation of Tensorial Components) as described in \cite{LEE2004945MITC3} for our FEA implementation, which is a linear shell model. The stiffness matrix is defined as $\mathbf{K}(d) = \cup_e (f_E(d_e) \cdot \mathbf{k}_0)$ following  \cite{CHI2021112739} with $\cup_e$ denoting the element assembly. This includes MITC3 shell element’s stiffness matrix $\mathbf{k}_0$, and output of Young’s modulus function $f_E(d_e)$, with the detailed function given in Sec.\ref{sec:blending} . The MITC3 element accounts for the thickness dimension and includes an additional internal node to capture the bending behavior.
Note that the boundary condition of fixing the nodes on the boundaries of a brace needs to be imposed for FEA. $V_0$ is the allowed maximal volume, $V_e$ is the volume of element $e$, and $\mathbf{F}_{\text{acting}}$ indicates the applied nodal force, and $d_e$ is the design variable of the element $e$. The discrete compliance on each element $e$ can be defined as $W_c = f_E(d_e)\cdot\mathbf{u}_e^{T}\cdot\mathbf{k}_0(t_i)\cdot\mathbf{u}_e$ with $\mathbf{u}_e$ being the nodal displacement of element $e$. 
We further employ $p$-norm method to approximate the maximum compliance across all the ${N_t}$ frames 
as given in Table \ref{table:loss}, Sec.~\ref{sec:lossFunc}. The reaction force can be computed by the displacement $\mathbf{U}(t_i)$ and the acting force $\mathbf{F}_{\text{acting}}(t_i)$ as
\begin{equation}\label{eq:reactionForce}
\mathbf{F}_{\text{reaction}}(t_i) = \mathbf{K}(d) \mathbf{U}(t_i) -\mathbf{F}_{\text{acting}}(t_i)
\end{equation}
This reaction force serves as the brace support and acts back to the body to update the joint angles $\phi$ and the joint contact forces $\mathbf{F}_{\text{JCF}}$ sequentially with embedded solvers of OpenSim \cite{opensim4352056}, where the brace reaction forces are initiated with small values and updated during optimization with automatic differentiation supported by  \cite{Falisse2019CasADiOpensim, CasADi}.

\subsubsection{Elastic Energy Based Segmentation}
\label{sec:SegMultiple}
We employ elastic energy density to estimate the degree of local deformation on the time-variant surfaces of human body. To alleviate the constrained movement caused by the brace wrapping, we aim to minimize the total elastic energy on the brace. The distribution of elastic energy can be commonly evaluated by the design function 
$d(\cdot)$ and the deformation of elements on a brace among different frames. Specifically, given $\mathbf{G}_{e}(t_i)$, the deformation gradient~\cite{sumner2004deformation} of the element $e$ at the time frame $t_i$ w.r.t. its rest shape, the elastic energy $W_e$ at every element can be evaluated with the classical compressible neo-Hookean model ~\cite{bonet2008nonlinear} as
\begin{equation}\label{eq:elastic_energy}
 W_e=\max_{\{ t_i \} } \left\{\frac{\mu_e}{2} (I_{e}(t_i)-2)-\mu_e \log A_e(t_i) + \frac{\lambda_e}{2} (\log A_e(t_i))^2 \right\}
\end{equation}
where $I_{e}=tr(\mathbf{C}_e)$ is the first invariant of the right Cauchy-Green tensor $\mathbf{C}_e = \mathbf{G}_{e}^T \mathbf{G}_{e}$, and $A_e = \det(\mathbf{G}_{e})$ is the relative area change. 

$\lambda_e = \nu \frac{f_E(d_e) }{(1 + \nu) (1 - 2 \nu)}$ and $\mu_e =  \frac {f_E(d_e)}{2 (1 + \nu) }$ are the Lam\'e’s first parameter and the shear modulus respectively, and $\nu$ is the Poisson's ratio. Both are influenced by the design function $d(\cdot)$. Note that the maximal elastic energy among all relevant frames is employed here.

By the distribution of elastic energy $W_e$ (see the right figure), we segment the entire design domain into the soft and firm regions by the self-tuning spectral clustering method~\cite{zelnik2004self} so that different types of microstructures can be assigned into these regions respectively (ref.~\cite{liu2023dynamic}). \begin{wrapfigure}[6]{r}{0.28\linewidth}\vspace{-8pt}
\begin{center}
\hspace{-25pt}\includegraphics[width=1.0\linewidth]{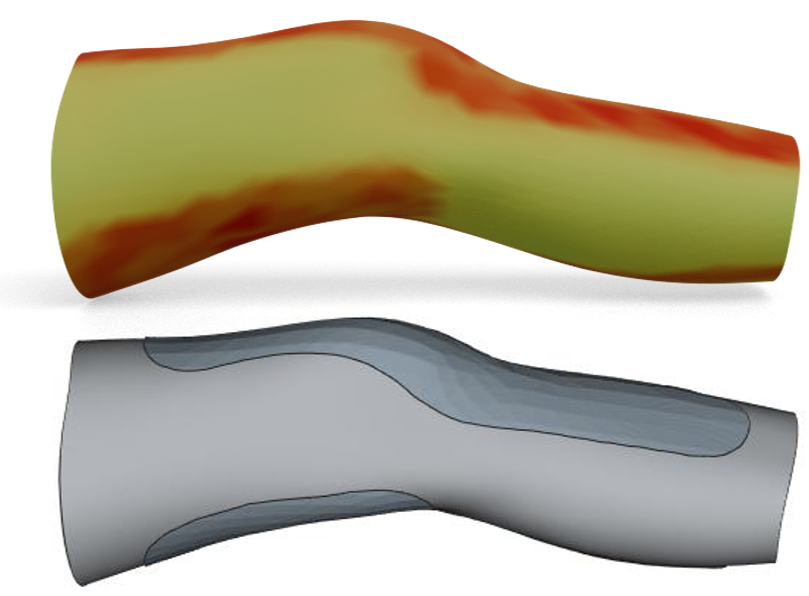}
\end{center}
\end{wrapfigure}When fabricating the braces with the same super-elastic material (i.e., silicone in our trials), we leverage the strength and stiffness of the microstructures type with high-stiffness in the region with low elastic energy and meanwhile also benefit from the flexible deformation characteristics of the type with low-stiffness in the region with high elastic energy. We introduce details of generating distributed microstructures with different stiffness in Sec.\ref{sec:FGM}
\section{Distributed Microstructures}\label{sec:FGM}
\subsection{Periodical formulation} 
\begin{figure}[t]
\setlength{\unitlength}{\linewidth}
  \begin{picture}(1,0.28)%
      \put(0,0){\includegraphics[width=\linewidth]{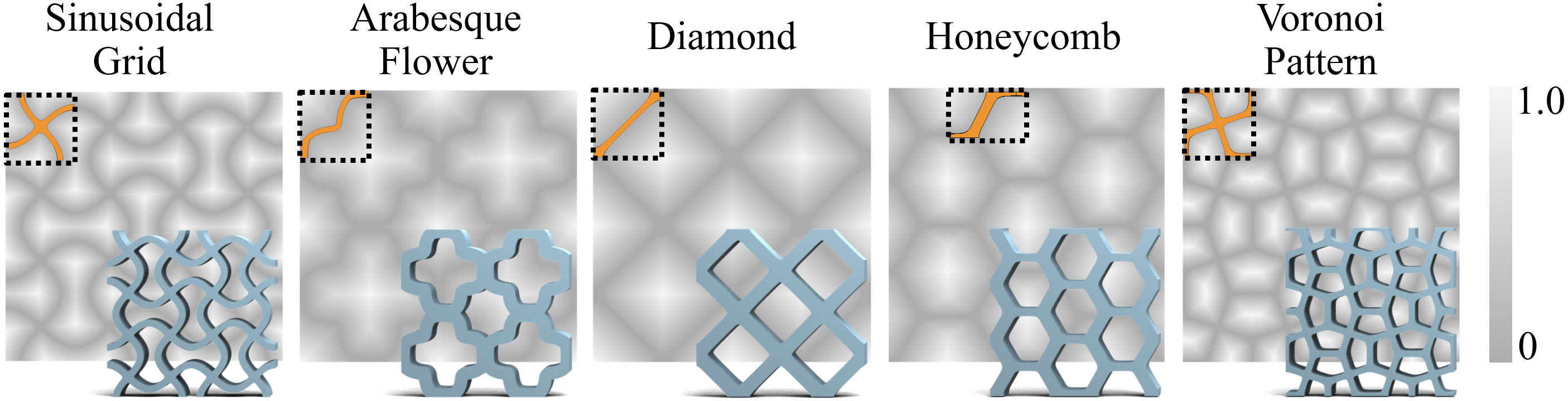}}%
    \small
  \end{picture}
\caption{
Different types of microstructures are studied in our research. We provide structures represented by an implicit surface $f_{cell}(\hat{u},\hat{v})$ in the $u,v$-domain, and the corresponding solids obtained by $f_{cell}(\hat{u},\hat{v}) \leq 0.2$ with a user-specified thickness are displayed in blue at the bottom right corner. The function $f_{cell}(\hat{u},\hat{v})$ is formulated as the distance field to the skeletons (highlighted by curves in orange). 
}\label{fig:FGMCell}
\end{figure}

Different types of microstructures can be represented with the help of an implicit function $f_{cell}(\hat{u},\hat{v})$ defined as the distance field to the skeletons in a unit parametric domain $(\hat{u},\hat{v}) \in [0,1] \times [0,1]$. The implicit solid of the distributed microstructures in the domain with design variable $d(u,v)$ can be formulated by a periodic function as
\begin{equation}
    f_{M}(d(u,v),u,v) = d(u,v) - f_{cell}(\hat{u},\hat{v}) 
\end{equation}
with $\hat{u}=\arccos(\cos {2 u\pi }/{s})/\pi$, $\hat{v}=\arccos(\cos {2 v\pi }/{s})/\pi$, and $s$ being the scale coefficient of an unit cell in the $u,v$-domain. The implicit solid of distributed microstructure can then be obtained as $\{(u,v) \, | \, f_{M}(u,v) \leq 0 \}$ -- see Fig.\ref{fig:FGMCell} for five different types of microstructures when $d(u,v)=0.2$. Some similar structures can be found in \citet{mechProp10.1145/3197517.3201278}. The value of scale coefficient $s=0.08$ is determined by experiments and employed in all examples in this paper. Note that the function of volume ratio $f_V(\cdot)$ for different types of microstructures is obtained by sampling different values of $d$ and evaluating the volumes of resultant implicit solids. The function $f_V(\cdot)$ is then approximated by a cubic polynomial.

\subsection{Mechanical property}\label{sec:MechProperty}
Young's modulus plays a vital role, it indicates a material's stiffness and elasticity~\cite{murugan2020mechanical}. Different types of microstructures possess Young's modulus with a broader range than those of conventional materials, which is ideal for designing prophylactic braces.

Young's modulus for microstructures can be approximated by homogenized elasticity tensors. We have evaluated Young's modulus of five types of microstructures shown in Fig.\ref{fig:FGMCell} by the method presented in~\citet{korner2014systematic}. Curves between Young's modulus and volume ratio, generated using 25 sample points and Least-Squares fitting with an approximation error less than 1e-4, are shown on the right. It can be observed that the \textit{Voronoi Pattern} structure contains the highest Young's modulus with the same volume ratio, whereas the \textit{Sinusoidal Grid} structure gives the lowest \begin{wrapfigure}[8]{r}{0.4\linewidth}\vspace{-10pt}
\begin{center}
\hspace{-15pt}\includegraphics[width=1.0\linewidth]{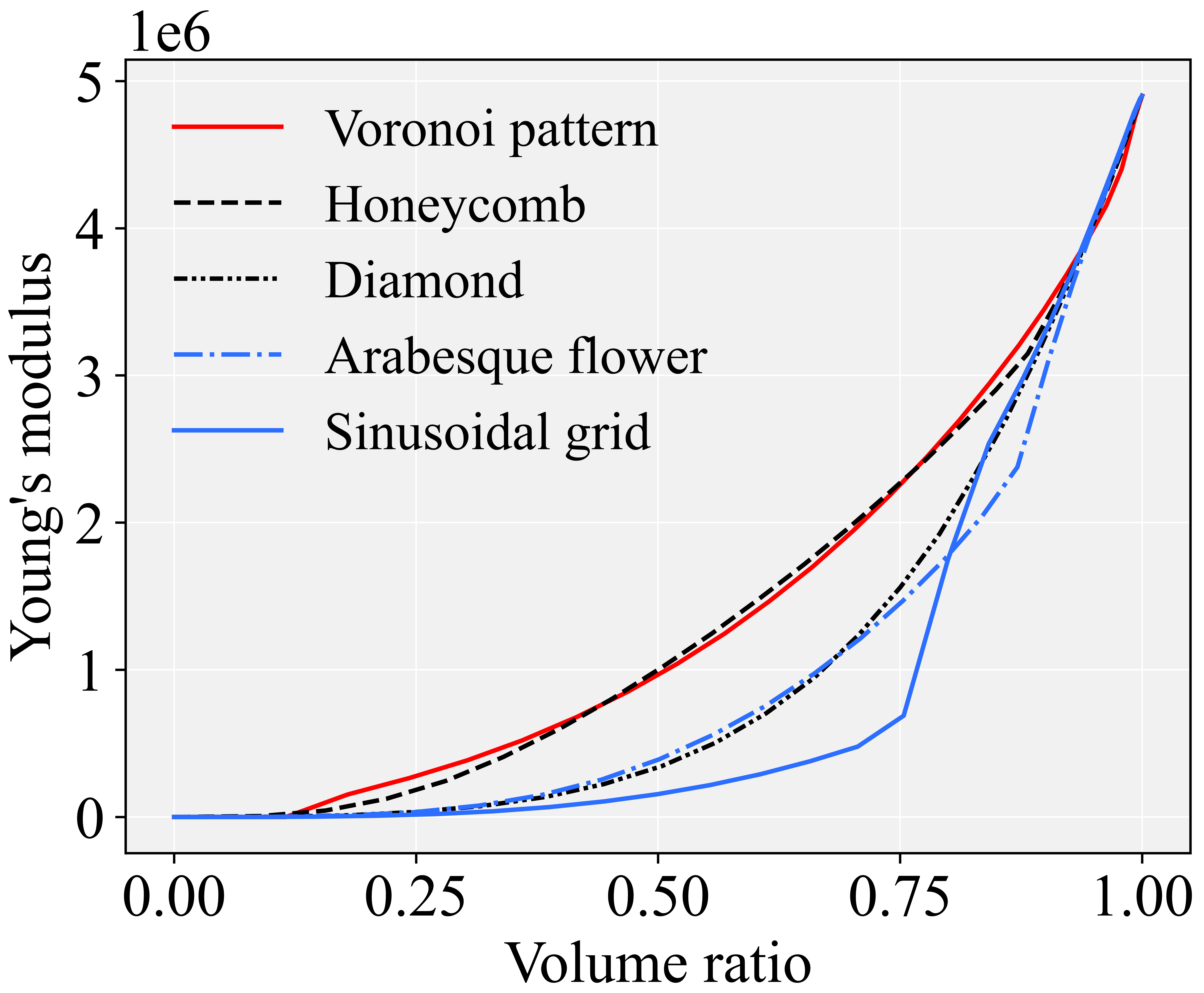}
\end{center}
\end{wrapfigure}Young's modulus. Following the strategy adopted in \citet{watts2019simple}, we use a cubic polynomial function to fit Young's modulus function $f_E(\cdot)$ depending on the design variable $d$. The resultant continuous and homogeneous Young's modulus functions for both the soft and the firm microstructures are then applied to conduct FEA in Eq.(\ref{eq:top}) for design optimization, where the volume constraint is controlled with the help of the function $f_V(\cdot)$. 

\begin{table}
  \caption{
  Losses for Our Optimization Pipeline
  }
\small
 \begin{tabular} {p{0.14\textwidth}@{\hspace{0.1cm}}l}
    \toprule
    Loss & Formulation\\
    \hline
    \hline
    Joint Stability &$\mathcal{L}_1 = \sum_{s\in S_{\text{preserve}}} (\Tilde{\phi_s}-\phi_s)^2 + \sum_{s\in S_{\text{prevent}}} \Tilde{\phi_s} ^2$\\
    Elastic energy density & $\mathcal{L}_2 = \sum_{e=1}^{N_l} W_e$\\
    Compliance &$\mathcal{L}_3 = \frac{1}{2} (\sum_{i=1}^{N_t} (\sum_{e=1}^{N_l} W_c(d_e, t_i) ) ^{p} )^{\frac{1}{p}}$\\
    Volume fraction&$\mathcal{L}_4 = ( \sum_e^{N_l} (f_{V}(d_e)\cdot V_e)/V_0 -1 )^2$\\
    Seam Continuity&$\mathcal{L}_5 = \sum_{\left(i,j\right) \in \mathcal{D}} \left( d_i - d_j\right)^2$\\
    \hline
    \hline
    Total Loss&$\mathcal{L}_{\text{total}}= 
        \omega_1 \mathcal{L}_1 + \omega_2 \mathcal{L}_2 + \omega_3 \mathcal{L}_3+ \omega_4 \mathcal{L}_4+ \omega_5 \mathcal{L}_5$ \\
  \bottomrule
\end{tabular}
\raggedright
\footnotesize
\label{table:loss}
\end{table}

\begin{table}
  \caption{List of symbols employed in the loss functions}
\small
  \begin{tabular}{p{0.8cm}p{7cm}}
    \toprule
    Term & Explanation \\
    \hline \hline
    $S_\text{preserve}$ & Set of DoFs that need to be preserved (e.g. sagittal angles) \\
    $S_\text{prevent}$ & Set of DoFs that need to be controlled (e.g. frontal angles) \\
    $\Tilde{\phi_s}$& Updated joint angle for $s$ given the reaction as defined in Eq.(\ref{eq:reactionForce}) \\
    $\phi_s$ & The target joint angle for $s$ \\
    $W_e$ & The elastic energy on an element $e$ as defined in Eq.(\ref{eq:elastic_energy})\\
    $p$& The parameter for $p$-norm method ($p=20$ in our trials)\\
    $\mathcal{D}$& The pairs of corresponding points on the cutting line\\
    $d_i$ & The value of design variable on the $i$-th node of the mesh $\mathcal{M}$\\
    $d_e$ & The design variable sampled at the center of an element $e$\\
  \bottomrule
\end{tabular}
\label{table:symbol}
\end{table}

\subsection{Blending dual microstructures} \label{sec:blending}
After segmenting the design domain of a brace into soft and firm regions with the help of elastic energy obtained by Eq.(\ref{eq:elastic_energy}), we generate a blending map $f_{seg}(u,v)$ by assigning $f_{seg}(u,v)=0$ for the soft region and $f_{seg}(u,v)=1$ for the firm region, and a scale (0,1) for the blended regions. To ensure a smooth transition, the boundary values of $f_{seg}(u,v)$ between soft and firm regions are smoothed. The implicit solid can be defined by blending as 
\begin{equation}
f_{\bar{M}}(d(u,v),u, v) = \begin{bmatrix}
  1-f_{seg}(u, v)\\
  f_{seg}(u, v)
\end{bmatrix}^T
\begin{bmatrix}
 f_{M,\text{soft}}(d(u,v),u, v)\\
f_{M,\text{firm}}(d(u,v),u, v)
\end{bmatrix} 
\end{equation}
where $f_{M,\text{soft}}(u, v)$ and $f_{M,\text{firm}}(u, v)$ are the functions of the soft and firm microstructures respectively. We blend the Young's modulus the similar way, with the function $f_{\bar{E}}=(1-f_{seg})f_{E,\text{soft}}+f_{seg}f_{E,\text{firm}}$, and  $f_{E,\text{soft}}$ and $f_{E,\text{firm}}$ are the functions of the soft and firm Young's modulus respectively. Such blending of Young’s modulus will introduce some error but in a very small surface region.

The right figure shows the comparison with (bottom) vs. without (top) the blending formulation. Note that the width of the blended \begin{wrapfigure}[8]{r}{0.25\linewidth}\vspace{-10pt}
\begin{center}
\hspace{-20pt}\includegraphics[width=1.0\linewidth]{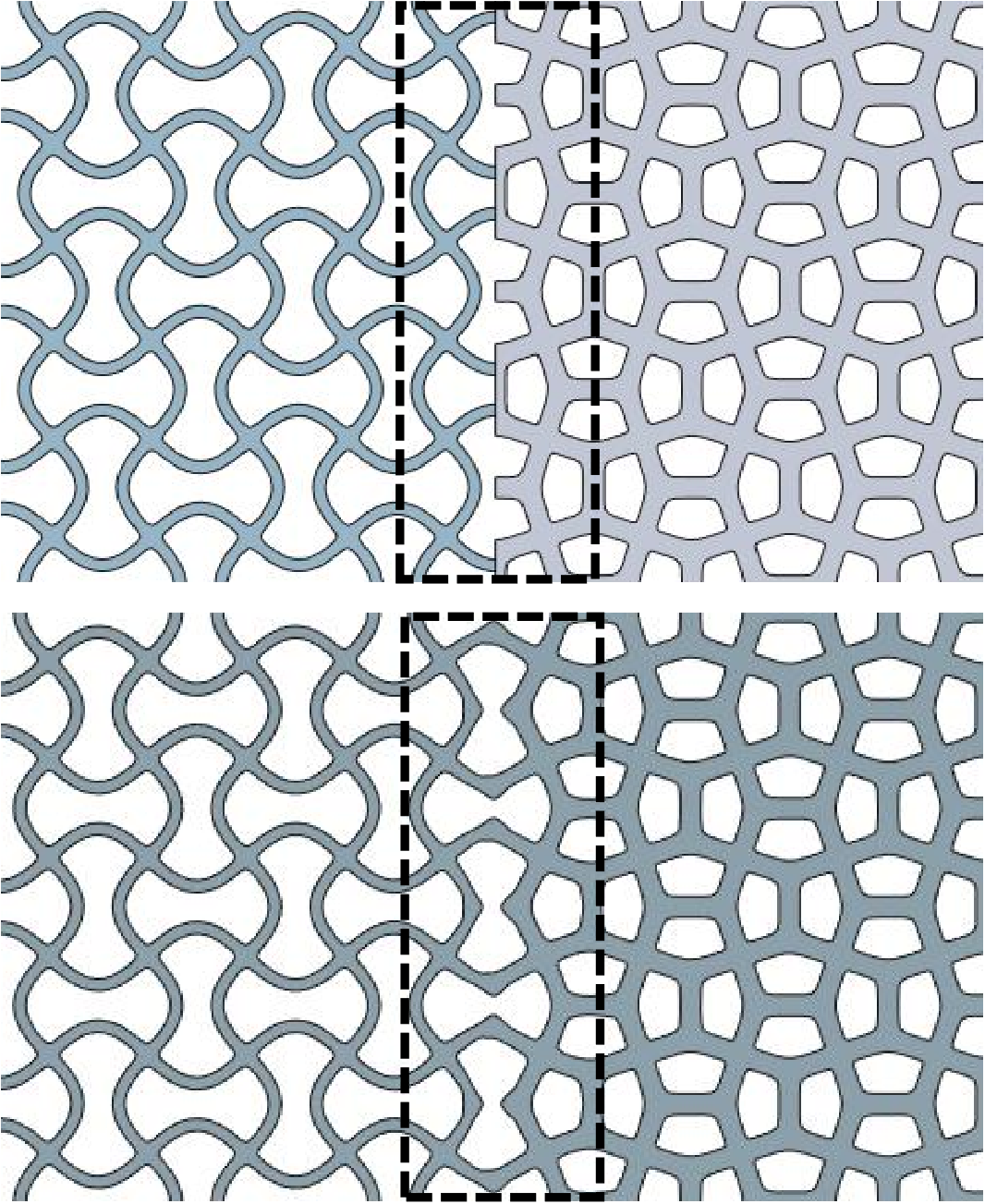}
\end{center}
\end{wrapfigure}region needs to be controlled due to the inaccuracy of Young's modulus of the blended microstructures. The microstructures as a distribution optimized on a brace can be seen in  Fig.~\ref{fig:pipeline}(d). The resultant solid can be obtained by first extracting a mesh with $f_{\bar{M}}(d(u,v),u,v) \leq d$ on the human body surface and then thickening the surface mesh into a solid~\cite{wang2013Offset, WANG2013321}.

\section{Loss Functions}
\label{sec:lossFunc}

Our optimization model aims to minimize the total elastic energy and compliance with reference to the acting force while considering the joint stability under the prescribed volume fraction. We define each loss term in Table~\ref{table:loss} where the definition of relevant symbols are listed in Table.~\ref{table:symbol}. The total loss is defined by combining all terms that are normalized by their initial loss values. 

Note that increasing the rigidity of the brace to prevent frontal angle may decrease the sagittal angle (restrict knee flexion), which leads to a trade-off of joint stability and mobility. Our design of loss function allows us to consider supportiveness and motion flexibility on a brace design concurrently. For practical implementation, seam continuity is considered by maintaining consistent design variables at both sides of a cutting line. Additionally, we incorporate total volume fracture loss for the lightweight nature of braces. Note that we only observed a very trivial discontinuity of $d(\cdot)$ across the cutting line in all examples if not applying this loss. However, the value of compliance loss $\mathcal{L}_3$ will be increased if the discontinuity is not penalized. Therefore, a very small weight $\omega_5=0.02$ is employed for the seam continuity loss while the other weights are chosen by experiments as $\omega_1 =0.5$, $\omega_2=0.8$, $\omega_3=0.6$ and $\omega_4=0.3$.

\section{Network Architecture}
\label{sec:network}
The aim behind introducing neural networks to present the distributed microstructures is to create a differentiable design function $d(\cdot)$ for a gradient-based optimizer. Specifically, $d(\cdot)$ is defined as a single layer neural-network:
\begin{equation}
d(w_1, w_2,...,w_{N_w}) = \sum_i^{N_w} w_i \psi_i(u,v),\quad \sum_i^{N_w} w_i=1 \\
\end{equation}
with ${N_w}$ being the number of neurons. The neural weights $\{w_i\}$ are the variables of the optimizer to influence terms defined in the loss function while serving as the differentiable representation for microstructures. In our implementation, the Gaussian covariance $\psi_i(u,v) = \exp \left(-\sigma^2((u-u_i)^2+(v-v_i)^2)\right)$ with $\sigma=4.0$ is chosen as the activation function because of its high flexibility and adaptability to non-linear data distributions. The centers of activation functions are obtained by uniformly sampling the $u,v$-domain into $N_w$ points. In addition, the projection is placed behind the single layer NN in order to keep the design function $d$ within the range $[0,1]$. For the sake of easy to converge, a single layer NN is employed in our implementation with ${N_w}=10,000$ neurons. The framework can be further extended to capture higher nonlinear variation in the design space by using multiple layers of neurons.

\begin{figure}[t]
\centering
\setlength{\unitlength}{\linewidth}
  \begin{picture}(1,1.46)%
      \put(0,0){\includegraphics[width=\linewidth]{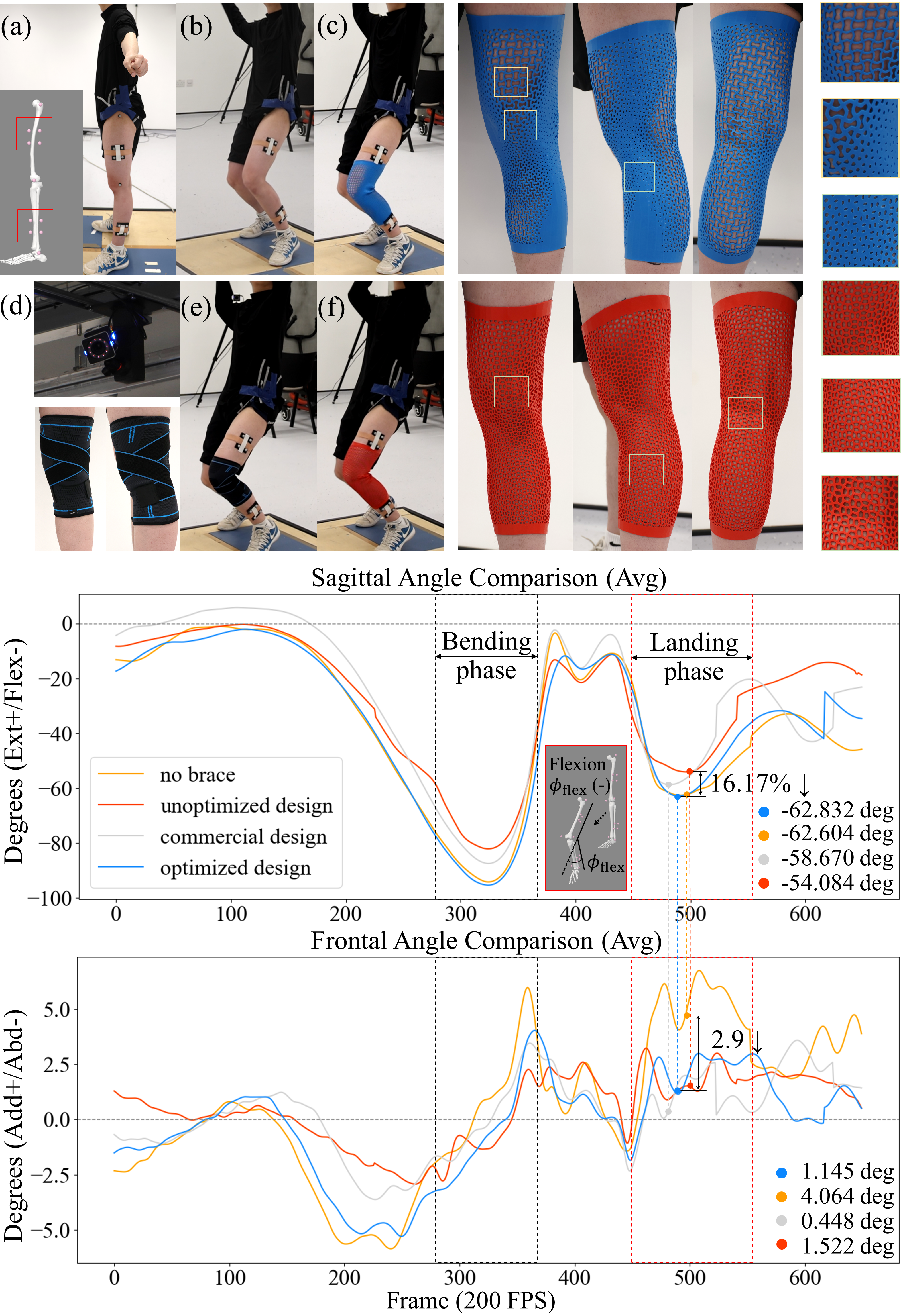}}%
    \small
  \end{picture}
\caption{Physical validation of prophylactic brace for the left knee. Left: (d) Motion capture system with twelve Vicon Vero 13 cameras. (a) Motion capture protocol for 13 markers placed on left lower extremity: 2 clusters on femur and tibia -- each with 4 markers (see red boxes), which are used to compute knee angles. Five more calibration markers located on left hip, lateral and medial of left knee and ankle to identify joint and body segments. The participant repeats the tennis serving motion 5 times for (b) unbraced, (c) wearing our optimized brace, (e) a commercial prophylactic brace and (f) an unoptimized brace with uniform `Voronoi Pattern' microstructure. Right: plot of joint angles for comparison, which are obtained from the average of synchronized curves for each case. Note the peak flexion is used for alignment and the second peak flexion is selected for comparing the angles for both planes at the landing moment, as this passive action can elevate the risk of joint injury, especially if the knee lacks adequate flexibility and remains extended upon landing. Our optimized design yields similar support as the uniform stiff patterns (reduced) and commercial braces while providing more flexibility within the sagittal plane, which is important to allow more bending upon contact with the ground.
}
\label{fig:evlaution_figonly}
\end{figure}

\begin{figure*}[t]
\centering
\setlength{\unitlength}{\linewidth}
  \begin{picture}(1,0.18)%
      \put(0,0){\includegraphics[width=\linewidth]{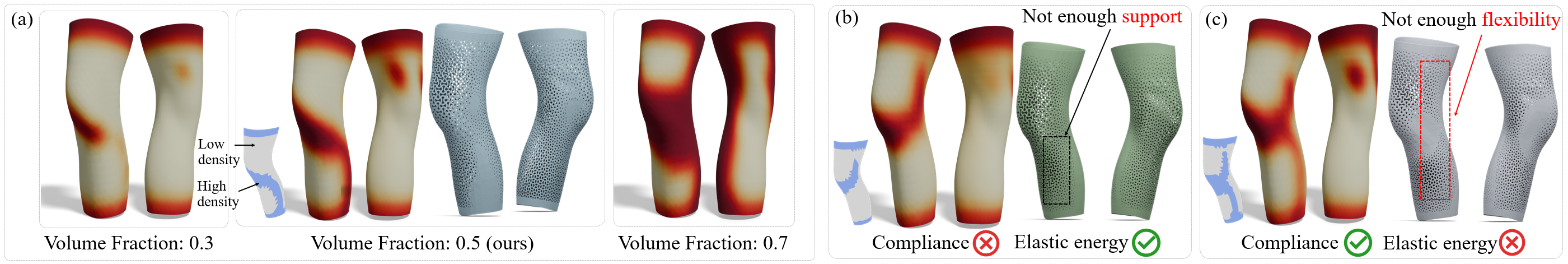}}
    \small
  \end{picture}
\caption{
Ablation study for the necessity of different loss terms: (a) The loss term of volume fraction $\mathcal{L}_4$ is employed to control the amount of materials used for the braces -- the volume fraction $0.5$ is employed for all other examples in this paper, (b) removing the compliance loss $\mathcal{L}_3$ will generate a brace with less support to prevent the adduction -- the upper \& lower loops of the brace are not connected by high-density regions, 
and (c) the flexibility of movement is not preserved without the loss of elastic energy density $\mathcal{L}_2$ 
which leads to a result with extra connection from the knee to the upper loop by a high-density region. 
}\label{fig:ablation}
\end{figure*}
\begin{figure}[t]
\centering
\setlength{\unitlength}{\linewidth}
  \begin{picture}(1,0.92)%
      \put(0,0){\includegraphics[width=\linewidth]{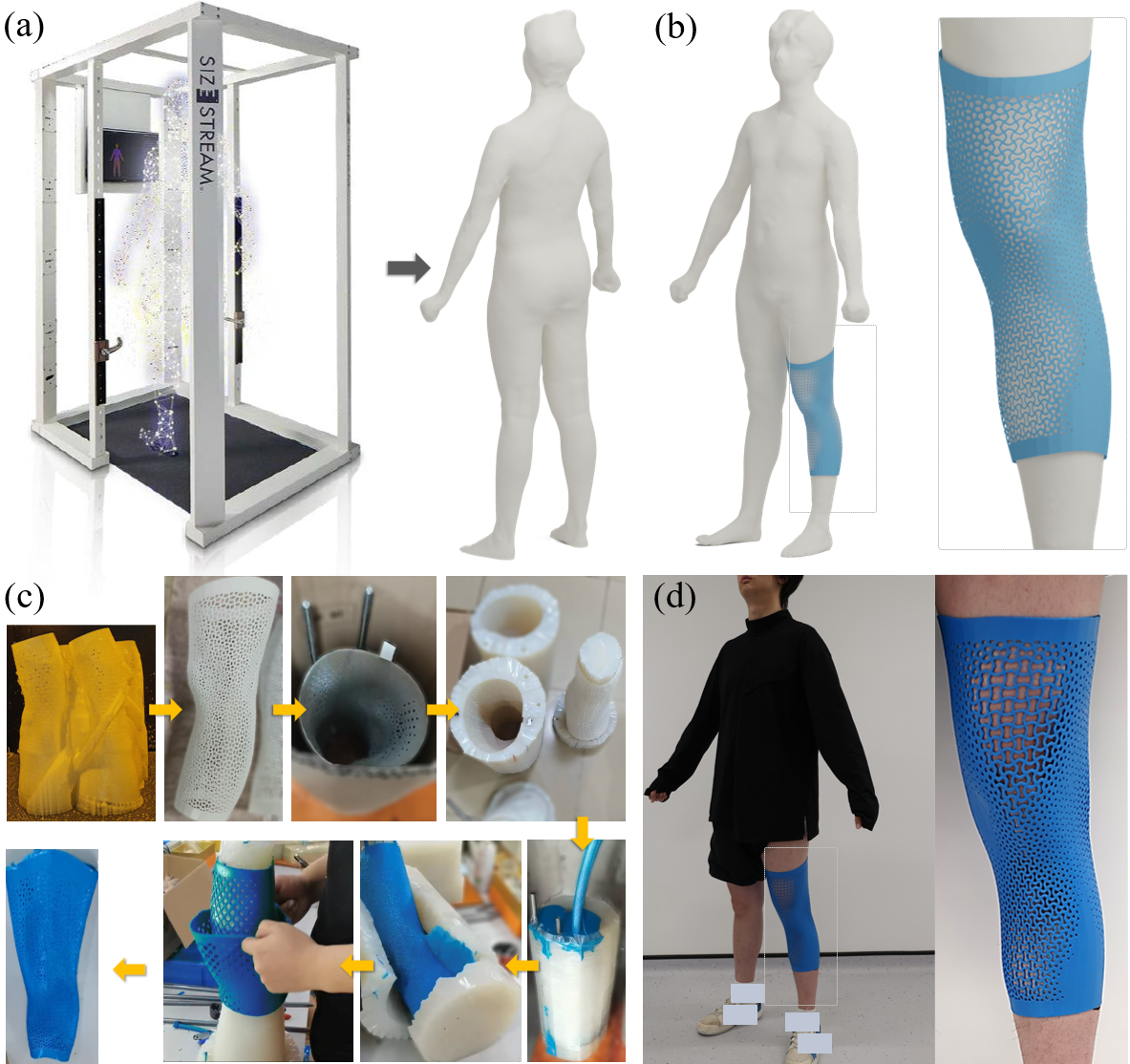}}%
    \small
  \end{picture}
\caption{To ensure the brace can be properly worn, we first capture the participant's body by a structure-light based scanner (online image resource accessed May 2024) and reconstruct the 3D surface (a). The optimized design of a knee brace is computed in a user-specified region (b). The designed brace with microstructures are generated by offsetting \cite{WANG2013321} and fabricated by Silicon rubber with 30A stiffness using a casting method with the 3D printed mold (c). (d) A physical try-on is provided with a zoom-in view on the right.
}\label{fig:fabrication}
\end{figure}

\section{Results and Validation}
We demonstrate our pipeline by analyzing and designing solutions tailored for different joint conditions and varied movement patterns. We fabricate the braces with Silicon rubber with 30A stiffness and verify their performance via physical try-on.

\subsection{Design for Prophylactic Braces and Ablation Study}
\begin{figure*}[t]
\centering
\setlength{\unitlength}{\linewidth}
  \begin{picture}(1,0.96)%
      \put(0,0){\includegraphics[width=\linewidth]{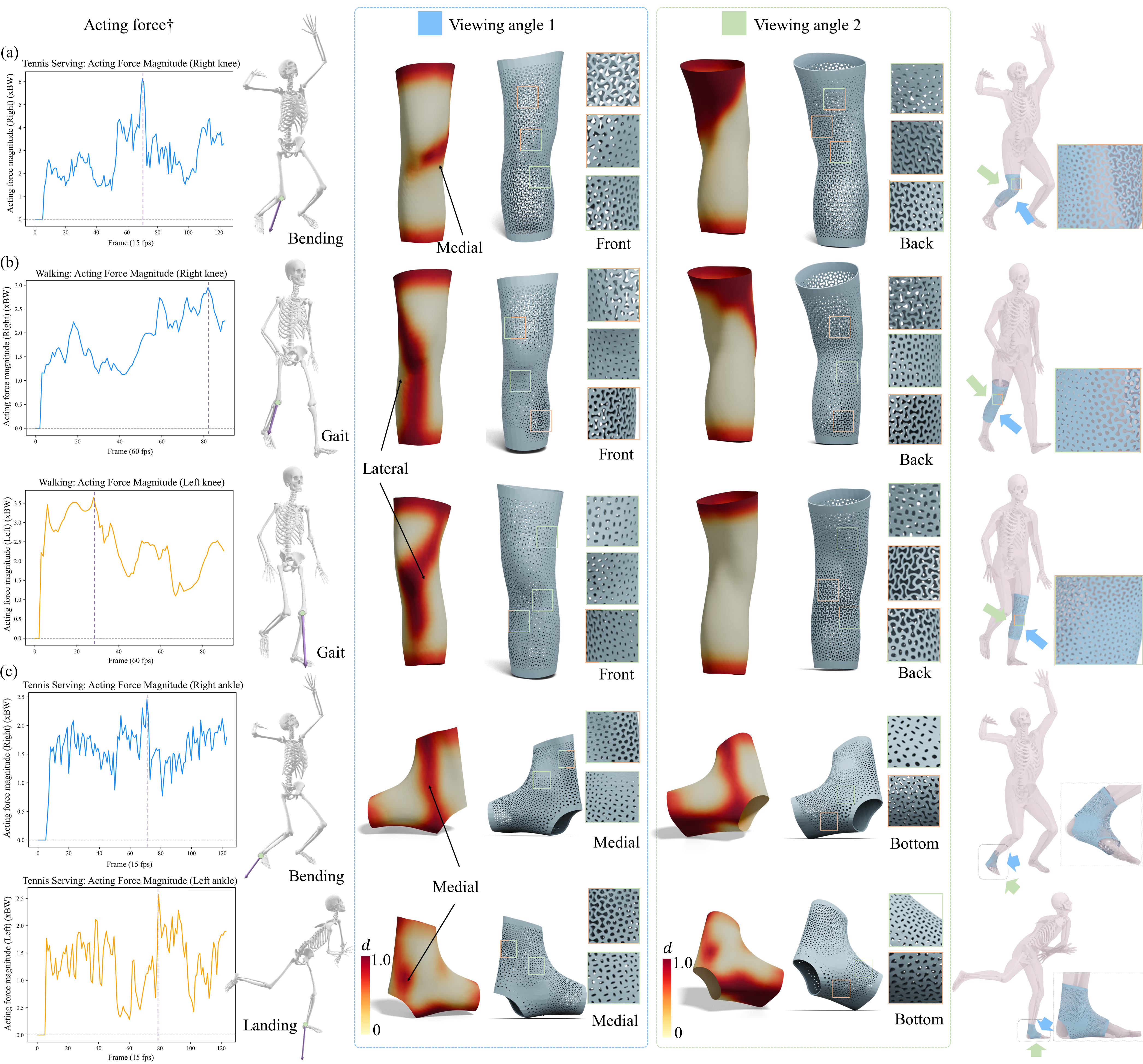}}%
    \small
  \end{picture}
\caption{
We present a collection of designed prophylactic braces, demonstrating the capability of the pipeline for designing braces for different joint dynamics and varied movement patterns. The columns are: (far-left) magnitude of time-variant acting force given by the joint contact forces and muscle forces; (second-far-left) peak acting forces on the joint; (middle) optimized design variable field and a detailed view of the braces' distributed microstructures from two different angles; (right) showcase of optimized design on the simulated human body of the same frame. Zoom-in views for firm (green) and soft (orange) regions are marked in boxes. (a) Compared with the left prophylactic knee brace for tennis motion (Fig.\ref{fig:teaser}), we yield different structure designs for the right side due to the asymmetry of human motion and different loads acting on the body. (b) Gait example: our optimizer generates more lateral support on the knee to help stabilize the joint and prevent lateral movement during walking. (c) Landing impact after tennis serve creates higher loads acting on the ankle joint. The designs for the left ankle specifically address the impact of bending and landing motions, which can significantly increase the risk of ankle injuries. The design result aims to provide protection for the medial sides of the foot that aims to mitigate eversion angles. (\textsuperscript{$\dagger$}The magnitude of the acting force are plotted.)
}\label{fig:designResults}
\end{figure*}

We first design a pair of knee prophylactic braces for tennis serving motion (see Fig.~\ref{fig:teaser} and Fig.~\ref{fig:designResults}(a)). During tennis serving that includes the phases of deep knee bending, jumping, and landing, a tennis layer can exert forces (normalized by the body weight) on their joints as shown in the force plots of Fig.~\ref{fig:teaser}. In the second example, we analyze the kinetics and design bilateral knee braces for gait (see Fig.~\ref{fig:designResults}(b)). Our optimizer effectively considers the asymmetry of the body and provides targeted support at the lateral of the knee for both sides. In the other example (see Fig.~\ref{fig:designResults}(c)), as different loads are applied to different joints, we also analyze the dynamics of the ankle for a tennis serving motion to generate customized ankle braces.

Furthermore, ablation studies have been conducted to illustrate the necessity of loss terms (see Fig.~\ref{fig:ablation}). The compliance loss $\mathcal{L}_3$ is important to provide enough support by controlling the frontal angle, and the elastic energy loss $\mathcal{L}_2$ can help effectively keep the flexibility in the sagittal plane. Removing the compliance loss $\mathcal{L}_3$ will generate a brace without direct connection (by high-density regions) between the upper and lower loop of the brace -- i.e., not enough support. On the other aspect, without the elastic energy loss $\mathcal{L}_2$, the design adds extra high-density regions, creating extra restriction to the movement. All these differences in the pattern topology of high-density regions are visualized in Fig.~\ref{fig:ablation} and quantative results are provided in Table.\ref{table:verification}.

\subsection{Physical Try-on} We fabricated the prophylactic braces for the left knee joint tailored to tennis-serving motion with silicone. Given the body surface of the scanned participant (Fig.~\ref{fig:fabrication}(a)), we obtain the shape parameters, $\beta$ of the SKEL ~\cite{keller2023skel} model by optimizing the parameters with a template human body to minimize the distance between two mesh surfaces. We conduct optimization on $(\beta, \mathbf{q})$, where the pose parameters $\mathbf{q}$ is from ~\citet{han2023groundlink}. Our optimized design conforms to the scanned body during physical try-on. The comparison of the scanned surface and fitted SKEL is shown in Fig.~\ref{fig:biomechModel}(c) and the detailed fabrication and try-on process can be found in Fig.~\ref{fig:fabrication}(c, d). 
\subsection{Verification}
\label{sec:verification}

\begin{table}
  \caption{Joint angle comparison for physical verification at bending and landing moments}
\small
  \begin{tabular}{p{0.2\textwidth}@{\hspace{0.18cm}}c@{\hspace{0.18cm}}c@{\hspace{0.18cm}}c@{\hspace{0.18cm}}c}
    \toprule
    Brace condition & Bending & Bending &  Landing & Landing  \\
    & Sagital$^{\dag}$ & Frontal$^{\ddag}$ & Sagital$^{\dag}$ & Frontal$^{\ddag}$ \\
    \midrule
    Unbraced $^{\ast}$  &0&0.825&0&16.519\\
    Unoptimized brace (Fig.\ref{fig:ablation}(b))&1.413&1.990&28.141&11.39\\
    Unoptimized brace (Fig.\ref{fig:ablation}(c))&78.049&4.045&33.091&7.537\\
    Unoptimized brace (Fig.\ref{fig:evlaution_figonly}(f))&144.148&1.782&72.579&2.315\\
    Commercial brace&44.788&\textbf{0.076}&15.478&\textbf{0.200}\\
    Optimized brace (ours)&\textbf{1.374}&0.526&\textbf{0.052}&1.310\\
  \bottomrule

\end{tabular}
\raggedright
\footnotesize
$^\dag$ Sagittal angle with extension (+) and flexion (-) is compared with $^{\ast}$. A sagittal angle close to unbraced condition suggests angle preservation.\newline
$^\ddag$ Frontal angle with adduction (+) and abduction (-) is compared with \textit{zero}. A frontal angle close to zero is indicative of more joint stability within frontal plane. \newline
\label{table:verification}
\end{table}

While the software solution for simulating the musculoskeletal model with the optimized brace model with distributed microstructures is challenging, we directly conduct physical experiments to validate the performance of optimized braces generated by our approach (see Fig.\ref{fig:evlaution_figonly}). We test the fabricated left knee brace for a tennis player and evaluate the joint angles from the mocap data during tennis serving, with protocols introduced in Fig.~\ref{fig:evlaution_figonly}(a). Our optimized design is compared against unbraced conditions, wearing unoptimized (uniformly firm) and commercial braces, as shown in Fig.~\ref{fig:evlaution_figonly}. We demonstrate that the braces with the optimized distribution of elasticity can help reduce the risk of injury by preserving the flexibility of movement within the sagittal plane (Column 1, 3 from Table.~\ref{table:verification}) while reducing the frontal angles (Column 2, 4 from Table.~\ref{table:verification}).

\section{Conclusion and Discussion}
We present a motion-driven neural optimizer that leverages biomechanical insights and physical formulation for the topology design of prophylactic braces tailored to human motions. The generated results are validated via physical tests, demonstrating tailor-made mechanical protection and preservation of the optimized braces.

Our approach has several limitations that should be acknowledged. First, the mesh density and quality used for FEA could impact the accuracy of the simulation results. Second, classifying a model into soft/firm regions introduces significant simplification, and more sophisticated methods using more microstructures can further improve the optimization results. Third, the shape of the brace is derived from a human body model created by fitting a SKEL model onto a scan, which may introduce errors. Additionally, the study simplifies the analysis by neglecting FEA on bone structures and not accounting for changes in muscle forces due to reaction forces generated by different brace elasticities. These simplifications and potential errors could influence the accuracy and prevent the generation of further optimized braces. Future work should address these limitations to improve the precision of the results.

\begin{acks}
The authors would like to thank the anonymous reviewers for their valuable comments. The authors also thank the participants for their time and commitment to the study. The physical experiments (motion capture) were conducted at University of Leeds and are approved by the Engineering and Physical Sciences Faculty Committee (LTELEC-001). The physical experiments (human body scanning) were carried out at the University of Manchester by the Apparel Design Engineering group, Department of Materials in accordance with the ethically approved process (Ref.\#: 2024-19436-34965, Study Title: Body Scanning at the University of Manchester 2024-2029 \cite{Gill2016/11}). The authors extend their thanks to their colleagues for the valuable discussions and assistance with the physical experiments. 

The project is supported by the chair professorship fund at the University of Manchester and UK Engineering and Physical Sciences Research Council (EPSRC) Fellowship Grant (Ref.\#: EP/X032213/1), the National Natural Science Foundation of China (No. 62172073, No. 62027826). This work was partially funded by a gift from Adobe.
This work uses aitviewer \cite{aitviewer_2022} and \citet{Liu_BlenderToolbox_2018} for visualization.
\end{acks}

\bibliographystyle{ACM-Reference-Format}
\bibliography{MotionSyn}

\appendix
\section{Notation table}
We provide a full list of notations employed in our main manuscript in Table.\ref{table:symbolfull}

\vspace{0.2cm}
\tablecaption{List of symbols employed in the paper}
\tablefirsthead{\toprule Term&\multicolumn{1}{l}{Explanation} \\ \midrule}
\tablehead{%
\multicolumn{2}{l}%
{{\bfseries  Continued from previous column}} \\
\toprule
Term&\multicolumn{1}{l}{Explanation}\\ \midrule}
\tabletail{%
\midrule \multicolumn{2}{r}{{Continued on next column}} \\ \midrule}
\tablelasttail{%
\\\midrule
\multicolumn{2}{r}{{Concluded}} \\ \bottomrule  
}

\small
\begin{xtabular}{p{1.5cm}p{6.5cm}}
$\beta$ & the shape parameters of SKEL model \cite{keller2023skel}\\
$\mathbf{q}$ & the pose parameters of SKEL model \cite{keller2023skel}\\
$e$ & the finite elements in FEA \\
$\mathbf{K}$ & the global stiffness matrix for FEA \\
$\mathbf{k}_0$ & the element’s stiffness matrix of MITC3 shell\\
$\mathbf{U}$ & the nodal displacement vector for FEA \\
$\mathbf{u}_e$ & the nodal displacement of an element $e$ \\
$\mathbf{F}_{\text{GRF}}$ & the ground reaction force \\
$\mathbf{F}_{\text{acting}}$ & the overall acting force \\
$\mathbf{F}_{\text{JCF}}$ & the joint contact forces \\
$\sum\mathbf{F}_{\text{mus}}$ & the sum of the muscle forces \\
$d(u,v)$ & NN based design function in the domain \\
$d_i$ & the value of design variable on the $i$-th node of the mesh $\mathcal{M}$\\
$d_e$ & the design variable sampled at the center of an element $e$\\
$f_E$ & the function of the Young’s modulus \\
$f_V$ & the function of the volume ratio\\
$V_e$ & the volume of an element $e$ \\
$N_l$ & the number of elements \\
$N_t$ & the number of frames \\
$t_i$ & the i-th time frame\\
$\phi$ & the joint angles of the human body \\
$W_c$ & the discrete compliance on an element $e$\\
$W_e$ & the elastic energy on an element $e$ as defined in Eq.~\ref{eq:elastic_energy}\\
$G_e$ & the deformation gradient of an element $e$ \\
$A_e$ & the relative area change of an element $e$ \\
$C_e$ & the right Cauchy-Green tensor of an element $e$ \\
$I_e$ & the first invariant of the right Cauchy-Green tensor of an element $e$ \\
$\lambda_e$ & the Lamé's first parameter of an element $e$\\
$\mu_e$ & the shear modulus of an element $e$\\
$\nu$ & the Poisson's ratio \\
$f_{cell}$ & the distance field to the skeletons in a unit parametric domain \\
$f_M$ & the implicit solid of the distributed microstructures in the domain\\
$f_{\bar{M}}$ & the blended implicit solid of the soft and firm microstructures in the domain\\
$f_{seg}$ & the blending map for recognizing soft and firm regions \\
$f_{M,soft}$ & the implicit solid of the soft microstructures in the domain\\
$f_{M,firm}$ & the implicit solid of the firm microstructures in the domain\\
$f_{\bar{E}}$ & the blended Young’s modulus function of the soft and firm microstructures in the domain\\
$f_{E,\text{soft}}$ & the Young’s modulus function of the soft microstructures\\
$f_{E,\text{firm}}$ & the Young’s modulus function of the firm microstructures\\
$N_w$ & the number of neurons \\
$w_i$ & the i-th neural weight \\
$\Phi_i$ & the Gaussian covariance \\
$\omega_i$ & the loss weight of the i-th loss term \\

$S_\text{preserve}$ & set of DoFs that need to be preserved (e.g. sagittal angles) \\
$S_\text{prevent}$ & set of DoFs that need to be controlled (e.g. frontal angles) \\
$\Tilde{\phi_s}$& the updated joint angle for $s$ given the reaction as defined in Eq.~\ref{eq:reactionForce} \\
$\phi_s$ & the target joint angle for $s$ \\
$p$& the parameter for $p$-norm method ($p=20$ in our trials)\\
$\mathcal{D}$& the pairs of corresponding points on the cutting line\\


\label{table:symbolfull}

\end{xtabular}%

\end{document}